# DEMO ion cyclotron heating: status of ITER-type antenna design

M. Usoltceva[1], V. Bobkov[1], H. Faugel[1], T. Franke[1,2], A. Kostic[1], R. Maggiora[3], D. Milanesio[3], V. Maquet[4], R. Ochoukov[1], W. Tierens[1], F. Zeus[1], W. Zhang[1]

[1]*Max-Planck-Institut für Plasmaphysik, 85748 Garching, Germany*
[2]*EUROfusion Power Plant Physics and Technology (PPPT) Department, Garching, Germany*
[3]*Politecnico di Torino, Italy*
[4]*Laboratory for Plasma Physics, ERM-KMS, 1000 Brussels, Belgium, TEC Partner*

The ITER ICRF system will gain in complexity relative to the existing systems on modern devices, and the same will hold true for DEMO. The accumulated experience can help greatly in designing an ICRF system for DEMO. In this paper the current status of the pre-conceptual design of the DEMO ICRF antenna and some related components is presented. While many aspects strongly resemble the ITER system, in some design solutions we had to take an alternative route to be able to adapt to DEMO specific. One of the key points is the toroidal antenna extent needed for the requested ICRF heating performance, achieved by splitting the antenna in halves, with appropriate installation. Modelling of the so far largest ICRF antenna in RAPLICASOL and associated challenges are presented. Calculation are benchmarked with TOPICA. Results of the analysis of the latest model and an outlook for future steps are given.

## 1. Introduction

An Ion Cyclotron Range of Frequency (ICRF) heating system is being designed for DEMO with nominal 50 MW injected power delivered by 3 in-port ICRF antennas. The ITER-type antenna has been under development since 2019 as the main concept option, capable of being installed in DEMO equatorial ports and profiting from the return of experience from ITER. Such an antenna is based on conventional multi-strap antennas [1], allowing for heating scenario optimization and impurities minimization via feeding variations [2]. The current work focuses on the optimization of in-vessel wave excitation, power coupling to plasma and interaction with plasma facing components, which depend fundamentally on the ICRF antenna design. This addresses the current priorities set for the ICRF system development within the EUROfusion Work Package Heating and Current Drive framework. Other ICRF functions like breakdown, wall conditioning, etc. are not being developed yet.

The antenna should be wider than the maximum toroidal port size of 1.08 m (from the DEMO limiter port design, to keep the Breeding Blanket (BB) structural integrity), to achieve large enough radiating area and low toroidal wavenumber $k_\parallel \approx$ 4-5 m$^{-1}$ needed for power coupling and core plasma heating including ion heating [3]. In this work, we develop the idea to split the antenna into two halves with the total area bigger than the port area and install the halves through the port one after another. Toroidal cuts in the front part (breeder zone) of the BB will be needed to fit the antenna.

The antenna optimization is still ongoing. The latest antenna model is presented here, which emerged as a result of numerous model iterations with variations of the number of straps, their size, septa shape, Faraday Screen design and the transmission line (TL) routing [4]. Results of 3D electromagnetic simulations of the latest model will be shown, with the following key achievements:

1) balanced TL voltages $U_{TL}$ on the feeders of all straps (maximized usage of available power);
2) successfully minimized averaged local parallel electric field $E_\parallel$ near the antenna boundaries with blankets (plasma-wall interaction mitigation);
3) optimized $k_\parallel$ spectrum.

The TL voltages should not only be balanced, but also lay within technologically acceptable range of values. As a reference, the ITER upper limit is 45 kV. To assess the level of $U_{TL}$ representative for expected experimental values in DEMO, 3D modelling with plasma was performed by using TOPICA [5] and RAPLICASOL [6]. It allowed estimating the typical discrepancies between the more realistic plasma cases and HFSS simulations with conducting load (sea water) routinely used to evaluate the 3 criteria listed above. All calculations shown in this paper were done at the frequency of 53 MHz, which is representative of the envisaged frequency range covering several heating schemes including the second harmonic tritium heating [3].

## 2. In-port antenna options

Three development approaches for the in-port antenna have been considered, based on the mounting procedure:

- Fig.1a: antenna with the smallest toroidal extent, fitting into the port – port-plug antenna. Such a solution might better comply with the engineering constraints, and the mounting concept can be taken over from ITER. However, such alternative will, for the same voltages imposed on the straps, couple less RF power. For the given port dimensions, an antenna with a maximum size of 1.04 m x 2.76 m (~2.9 m$^2$) is possible.
- The second approach in Fig. 1b is the antenna divided in two Poloidally Arranged Halves (PAH) with the additional toroidal cut into the breeding zone of BB. The PAH approach allows the largest possible area

---

with the same port dimensions; it could reach 2.17 m x 2.12 m (~4.6 m$^2$). This antenna is also easier to model, because a poloidal half can be used successfully for the optimization described in Section 3. However, because of the largest possible impact on the BB structure and very difficult installation procedure shown in [4], which involves rotation of the halves, this option was discarded.
- The same as the second option, the third one (Fig. 1c) requires a BB cut. It is divided in halves in a different way: Toroidally Arranged Halves (TAH). This approach has the maximum possible antenna dimensions of 1.54 m x 2.76 m (~4.3 m$^2$), toroidally narrower than the PAH antenna, but taking advantage of the full poloidal extent. It slightly reduces the impact on the BB compared to the PAH antenna. Same as PAH, this solution has a significantly different mounting procedure compared to the ITER port-plug antenna. Such a procedure has been elaborated as shown in Fig. 2. The moving structures allow a consecutive installation of the halves, taking into account the fact that the second half can be moved in only after the first one is brought fully inside the torus, therefore not blocking the way. The TAH antenna is the main direction of the recent design work.

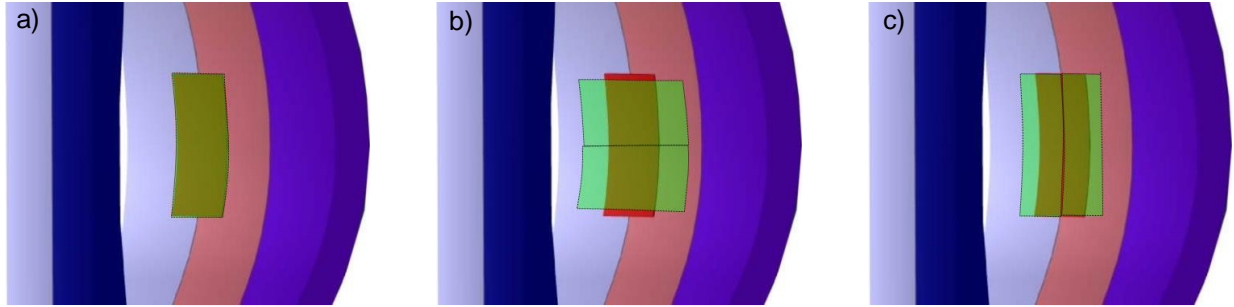

Figure 1. Mounting procedure approaches: a) port-plug antenna; b) PAH antenna; c) TAH antenna. The antenna is in green (semi-transparent), through it a red silhouette of the port can be seen. DEMO blanket sectors are in light-blue, pink and blue.

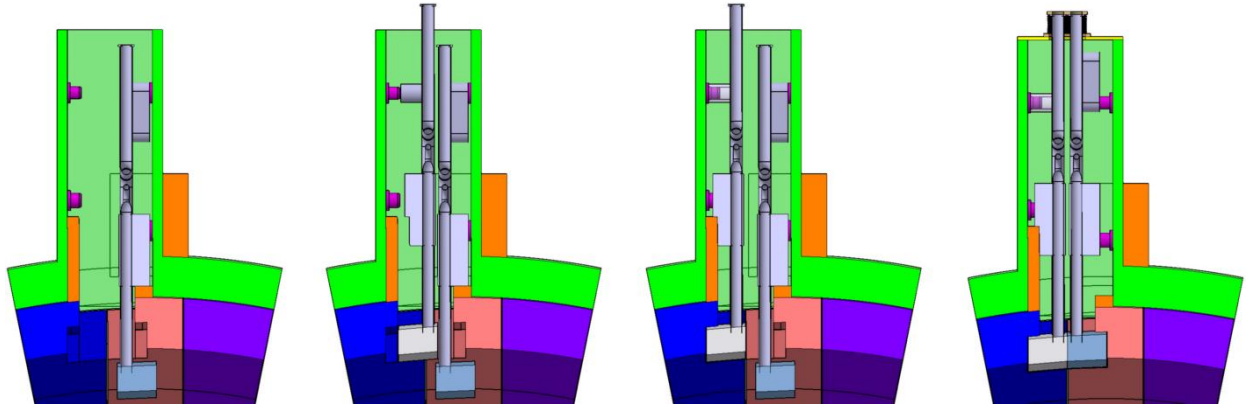

Figure 2. From left to right: sequence of TAH antenna mounting into the DEMO equatorial port, top view.

## 3. Antenna optimization

The coupled power depends significantly on the antenna design: (i) by possible voltage or current limitations in the straps and other components in the antenna system, and (ii) by the dominant $k_\parallel$ in the antenna spectrum. In general, smaller antennas will require larger voltages for the same power capability, because of the higher inductance of narrower straps. Also, for the same relative phasing of the straps $\Delta\varphi$, a smaller inter-strap distance $S_z$ will increase the dominant wavenumber $k_\parallel$ value, as follows from the relation $k_\parallel = \Delta\varphi/S_z$.

The RF performance of various antenna geometries is evaluated as follows:

a. Scan of P$_{central}$/P$_{outer}$ (ratio of the power coupled by central straps and the power coupled by outer straps) and the phasing in the proximity of the default dipole [0 π π 0] phasing provides an optimum – minimum of the parallel RF electric field $E_\parallel$ on the antenna sides.

b. For the found optimum feeding, the TL voltages are calculated for 16 MW coupled power per antenna (~50 MW in total from 3 antennas).

c. The $k_\parallel$ spectrum for the optimum feeding is analyzed. It should not only have the main component close to 4-5 m$^{-1}$, but also have the next peaks minimized, in order to ensure that most of the power goes into the main component.

The results of such evaluation can be seen in Fig. 3 for the latest antenna model (maximum size TAH antenna). The optimum yields P$_{central} \approx$ P$_{outer}$, which is good from the technical point of view. It is considerably shifted in phase from the default dipole, because of the additional electrical length of the outer strap feeding (details on it below). This shift corresponds to the phase on the feeding ports. The straps current still stays close to a dipole. The $E_\parallel$ level is quite well optimized, but this is for calculations with poloidal phasing 0°. If another phasing, like π/2, has to be used due to the need to employ 3 dB couplers, the $E_\parallel$ minimization has to be reassessed. The next step of the performed calculations showed well-balanced TL voltages with the maximum of 35 kV. This level is calculated for a sea water load located at 4.5 cm from the antenna after a vacuum layer, and corresponds to acceptable level in the plasma case, as will be further shown in Section 4. The last crucial result is favorable $k_\parallel$ spectrum, with small side peaks and the target dominant $k_\parallel$ value achieved.

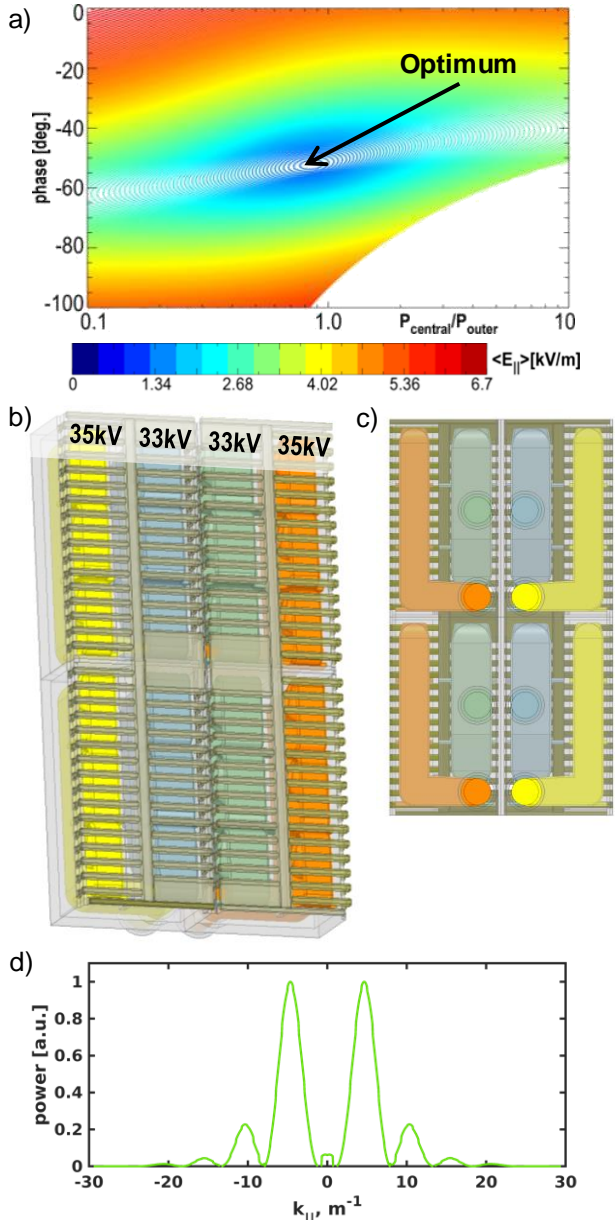

Figure 3. Analysis of TAH antenna model: a) optimum feeding, b) front view of antenna and TL voltages, c) rear view of antenna, d) $k_\parallel$ spectrum.

The parameters of this quite successful latest model are partially similar to ITER but in some aspects distinct. As in ITER, there are 8 straps (2 rows of 4), each strap split in triplets to maximize the power radiation. The Faraday Screen is also similar to that of ITER antenna. The septa shape was found to be a useful variable in the optimization. The best performance was achieved with central septum shorter than the box depth, which allows better cross-talk between the central straps, and with "shaped" outer septa (see [4]). The radial box depth is 0.5 m, corresponding to the allowed protrusion up to the BB edge. The coaxial transmission lines have the outer cylinder size of 9' = 229 mm; the inner size of 186 mm results in 12.5 Ω impedance. The transmission line routing is peculiar; the outer straps feeders were shifted as close as possible to the centre (by using strip-lines), as well as the inner straps (slightly), see Fig. 3c and [4]. In fact, their size and position define the toroidal size limit of an antenna half that can pass through the port (Fig. 2).

Crucial studies of neutron transport and activation have been initiated and results are expected in autumn 2020. In Fig. 4 one can see in orange an additional permanent neutron shield to protect the vacuum vessel. A removable neutron shield plug was added inside the port dedicated for ICRF antenna (not seen on the picture).

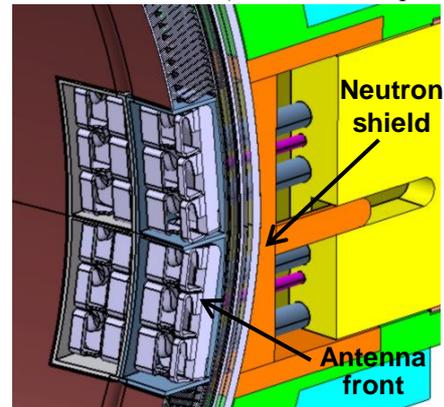

Figure 4. View from the DEMO vessel to the ICRF antenna and neutron shielding elements seen on the cut.

## 4. Benchmark with plasma

More sophisticated modelling was carried out with plasma, to benchmark it with the sea water HFSS simulations. In TOPICA and RAPLICASOL, magnetized plasma models are used to predict realistic values of the antenna loading. The plasma density was taken from the "ITER2010low" profile (as in [4]). The plasma starts at 105 mm from the antenna after a vacuum layer, which is introduced to replace the low-density plasma and avoid numerical difficulties in that region, caused by the need to resolve very fine slow wave structure and its interaction with the Lower Hybrid resonance layer.

This antenna is the largest of all ICRF antennas simulated so far in RAPLICASOL. The current study brought this tool to the edge of its capabilities on a cluster with 1 TB RAM: the well-resolved, convergence-tested final model needed 975 GB. The simulation of such large models, typical for DEMO and ITER antennas, became possible due to an extensive multi-step optimization work in COMSOL. Firstly, the size of the simulation domain had to be minimized. A balance was found when the plasma radial extent was large enough to include density ($9*10^{19}$ m$^{-3}$) with low enough wavelength (0.117 m) and it allowed having a relatively short Perfectly Matched Layer (PML), see Fig. 5. If the density was cut at a lower value, the plasma domain would be shorter, but due to much larger wavelength in lower density plasma the PML size would have to be few times larger, so the total model size would grow. Secondly, the mesh was optimized. It has appeared profitable to adjust the (tetrahedral) mesh element size for each domain separately. Besides, the relative scale is a useful feature which allows making the size of the element different in different directions. In our case the finest resolution in plasma is needed in the radial direction, so the element size in the two other directions can be few times larger. As an example, the plasma domain and the PML behind it which absorbs most of the power had the radial element sizes between $l_{min} = 0.008$ m and $l_{max} = 0.014$ m and relative scale $s$ up to 4. Meanwhile, the antenna volume had $l_{min} = 0.008$ m, $l_{max} = 0.08$ m and a rather low $s \sim 2$ (lower $s$ to represent antenna details). In the end, the total mesh had $2.5 \cdot 10^6$ elements and the model had $16.8 \cdot 10^6$ degrees of freedom. Quadratic discretization was used.

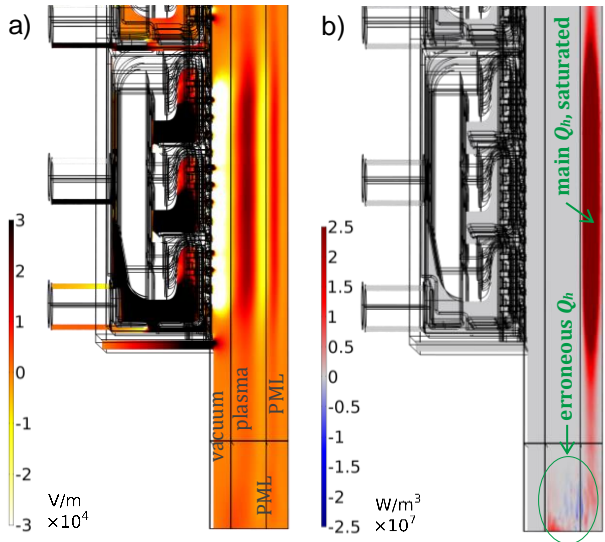

Figure 5. RAPLICASOL model, lower antenna half, side view on a cross-section in the middle of a powered strap: a) poloidal E-field, b) power dissipation density $Q_h$. The cross-section is centered at the powered strap.

The correct performance of the PML and plasma was proven according to 3 criteria:

- The wave pattern in plasma matches the analytical solution of cold plasma dispersion relation.
- The power dissipation locations in PMLs correspond to the wave propagation pattern and have their maxima well before the end of the PML, to ensure correct PML functioning.
- Quantitatively, the PML performance is evaluated from the integral $P$ of power dissipation density $Q_h$ and the integral $P'$ of its absolute value $|Q_h|$ in all model domains. The negative absorbed power $P^- = \frac{P - P'}{2}$ (unphysical incorrect PML behavior) should be much smaller than the positive one $P^+ = P - P^-$. In our model, the maximum $P^-$ was less than 1.5 % of $P^+$ for all feeding ports in the port scan.

These numerical errors were partially caused by the necessity to have an interface between vacuum and PMLs. The PML was adjusted to damp a wave coming from plasma, not from vacuum, so the field from the vacuum side caused some minor numerical errors (Fig. 5). The maximum power dissipation density in the problematic zones was on the order of 30 times less than the power dissipation density in the main absorption zone right in front of the antenna. It could be further reduced if the plasma domain was enlarged poloidally and toroidally, so that the evanescent field in vacuum could decay stronger and would cause smaller errors. In fact, the stronger decay in the toroidal direction led to the error there being an order of magnitude less than in the poloidal direction.

The values of TL voltages (at the same optimum feeding as in HFSS) calculated by RAPLICASOL match well the result of TOPICA: 48.0-43.8-43.8-48.0 kV and 47.9-43.8-43.8-47.9 kV, respectively. Similar to [4], ~35 % of voltage underestimation by HFSS is obtained. The ratio of the voltages at the inner and outer straps is reliably reproduced by all codes; the maximum voltage is close to the 45 kV limit. It is important to note that the voltage of 45 kV in the transmission line corresponds approximately to 41 kV on the straps, the latter being the limit used in ANTITER II [7]. With the same "ITER2010low" profile, both ANTITER II and the 3D modelling show that it is possible to couple power on the order of 50 MW from 3 antennas, if the antenna area is about 4 m$^2$.

## 5. Conclusions and outlook

The latest advancements of the development of ITER-type ICRF antenna for DEMO are shown, with the focus on RF performance and mechanical design. The optimization of a large number of design aspects allowed achieving a balanced antenna feeding and a favourable spectrum, while incorporating ITER experience on mechanical limitations. Advancements in RAPLICASOL allowed its successful simulation of the DEMO antenna.

Plasma kinetic profiles provided by the DEMO team will be used instead of "ITER2010low". The future work on the antenna itself will include a comparison of triplets and quadruplets (which could be better because the latest model has quite long straps), possible modifications of the box depth (partially) due to the interface with the BB on the sides, more detailed RF analysis including local electric fields and other optimization of the details.

## Acknowledgments

This work has been carried out within the framework of the EUROfusion Consortium and has received funding from the Euratom research and training programme 2014-2018 and 2019-2020 under grant agreement number 633053. The views and opinions expressed herein do not necessarily reflect those of the European Commission.

## References

[1] F. Durodié et al., AIP Conf. Proceedings 1580 (2014) 362
[2] V. Bobkov et al., Nucl. Mater. Energy 18 (2019) 131–140
[3] D. Van Eester et al., Nucl. Fusion 59 (2019) 106051
[4] V. Bobkov et al.., "Development of ITER-type DEMO ICRF antenna", submitted to Nucl. Fusion
[5] V. Lancellotti et al., Nucl. Fusion 46 (2006) S476
[6] J. Jacquot et al., Plasma Phys. Control. Fusion 55 (2013) 115004
[7] V. Maquet et al., 31st Symposium on Fusion Technology Proceedings (2020)